\documentclass[bibyear]{aa}  
\usepackage{txfonts}
\usepackage{url}
\usepackage{enumitem}

\usepackage{graphicx}

\def\spose#1{\hbox to 0pt{#1\hss}}
\newcommand\lsim{\mathrel{\spose{\lower 3pt\hbox{$\mathchar"218$}}
     \raise 2.0pt\hbox{$\mathchar"13C$}}}
\newcommand\gsim{\mathrel{\spose{\lower 3pt\hbox{$\mathchar"218$}}
     \raise 2.0pt\hbox{$\mathchar"13E$}}}

\def\ltsima{$\; \buildrel < \over \sim \;$}
\def\lsim{\lower.5ex\hbox{\ltsima}}
\def\gtsima{$\; \buildrel > \over \sim \;$}
\def\gsim{\lower.5ex\hbox{\gtsima}}

\def\apjl{ApJL}
\def\apjs{ApJSS}

\def\aap{A\&A}

\usepackage[normalem]{ulem}

\begin{document}

\title{Coherent curvature radiation and fast radio bursts }
\titlerunning{Coherent curvature radiation and FRB}
\authorrunning{Ghisellini \& Locatelli}
\author{Gabriele~Ghisellini\inst{1}\thanks{E--mail: gabriele.ghisellini@brera.inaf.it},
Nicola Locatelli\inst{1,2,3} \\
}
\institute{$^1$  INAF -- Osservatorio Astronomico di Brera, Via Bianchi 46, I--23807 Merate, Italy \\
$^2$  Univ. di Milano Bicocca, Dip. di Fisica G. Occhialini, Piazza della Scienza 3, I--20126 Milano, Italy \\
$^3$  Univ. di Bologna, Via Zamboni, 33 - 40126 Bologna, Italy    }

\abstract{
Fast radio bursts are extragalactic radio transient events lasting a few milliseconds with
a $\sim$Jy flux at $\sim$1 GHz.
We propose that these properties suggest a neutron star progenitor, and focus on 
coherent curvature radiation as the radiation mechanism.
We study for which sets of parameters the emission can fulfil the observational constraints.
Even if the emission is coherent, we find that self--absorption can limit the 
produced luminosities 
at low radio frequencies and that an efficient re--acceleration process
is needed to balance the dramatic energy losses of the emitting particles.
Self--absorption limits the luminosities at low radio frequency, while 
coherence favours steep optically thin spectra. 
Furthermore, the magnetic geometry must have a high degree of order to obtain coherent curvature emission. 
Particles emit photons along their velocity vectors, thereby greatly reducing the inverse
Compton  mechanism.
In this case we predict that fast radio bursts emit most of their luminosities
in the radio band  and have no strong counterpart in any other frequency bands. 
}
\maketitle

\section{Introduction}
Fast radio bursts (FRB) are ultrafast radio transients that are typically a millisecond
in duration with a flux at the Jy level at $\sim$1 GHz.
Their extragalactic origin is suggested by the measured dispersion measure (DM)
exceeding the Galactic value.
Recently, FRB 121102 has been associated with a galaxy at a redshift $z=0.19$,
roughly confirming the distance estimated with the observed DM (Chatterjee et al. 2017;
Tendulkar et al. 2017).


The extragalactic nature of FRBs implies luminosities around $10^{43}$ erg s$^{-1}$
and energetics of the order of $10^{40}$ erg.
The large flux, short duration, and cosmological distances also
imply a huge brightness temperature of the order of $T_{\rm B}\sim 10^{34}$--$10^{37}$ K.
This in turn requires a coherent radiation process.

Many models have already been proposed to explain the origin of FRBs\footnote{There is a strict
analogy with the pre--{\it Swift} era of gamma ray bursts, when the uncertainty about
their origin allowed hundreds of proposed scenarios for their origin}.
Before the discovery of the host galaxy of FRB 121102 even the Galactic scenarios were
believed to be feasible, although by a minority of the scientific community (e.g.
Loeb, Shvartzvald \& Maoz 2014;  Maoz et al. 2015).
Now this idea is disfavoured, but not completely discarded, since the repeating 
behaviour of this FRB remains unique. This makes the existence of {\it two} classes
of FRBs still possible, which is analogous to the early times of gamma ray bursts and 
soft gamma ray repeaters.

Progenitor theories of extragalactic FRBs include the merging of compact objects 
(neutron stars: Totani 2013, white dwarfs: Kashiyama et al. 2013),
flares from magnetars and soft gamma--ray repeaters
(Popov \& Postnov 2010; Thornton et al. 2013; Lyubarsky 2014, Beloborodov 2017), 
giant pulses from pulsars (Cordes \& Wasserman 2015), 
the collapse of supramassive neutron stars (Zhang 2014; Falcke \& Rezzolla 2014),
and dark matter induced collapse of neutron stars (Fuller \& Ott 2015).
There are even exotic ideas, such as FRBs explained by the propulsion of extragalactic sails
by alien civilizations (Lingam \& Loeb 2017) 
or the result of lightning in pulsars (Katz 2017).





        



A huge brightness temperature is probably the most important property of FRBs.
There are two mechanisms that can result in coherence: the first is a maser and the second
is bunches of particles contained in a region of a wavelength size, 
accelerated synchronously in such a way that their electric fields add up in phase.
In this second case the produced radiation depends on the square of the number
of particles in the bunch times the number of bunches.
In the maser case, particles are not required to be contained in a small volume, 
since the radiation is produced by stimulated emission and is emitted in phase and 
in the same direction of the incoming photons  by construction.
There are two kinds of masers. 
The usual type uses different, discrete energy levels of which one is metastable.
Free radiative transition from this level is somewhat inhibited, allowing
stimulated emission to be effective. 
This means that a maser requires a {\it population inversion} (high energy levels
more populated than low energy levels).
The second type of maser is associated with free electrons with no
discrete energy levels and there is no need for population inversion.
This type of maser occurs when a photon $h\nu$ can trigger stimulated emission with
a greater probability than true absorption.
This occurs rarely, but it does occur with some special configuration,
such as when electrons of the same pitch angle and energy emit,
by synchrotron, some photons outside the beaming cone of angle $1/\gamma$
; that is these electrons emit at relatively large angles with respect to their
instantaneous velocity (see e.g. Ghisellini \& Svensson 1991).

In Ghisellini (2017) we investigated whether a synchrotron maser can be
at the origin of observed radiation of FRBs.
The result was that it is possible, but as long as the magnetic field
is of the order of $B\sim$10--100 G if the emitters are electrons,
and $B\sim 10^4$-$10^5$ G if the emitting particles are protons.
These relatively weak values of the magnetic field are required
to produce radiation in the GHz radio band.
These values of $B$ are appropriate for main sequence stars (if the emitters
are electrons) and white dwarf (if protons), which would make it difficult to explain
large super--Eddington luminosities on a millisecond timescale.
We therefore think that the proposed synchrotron maser would be a valid model
if FRBs were Galactic (i.e. with a factor $\sim 10^{12}$ less energy and luminosity),
but it is less likely if all of these are extragalactic.

We (Locatelli \& Ghisellini 2017, herafter LG17) then explored the possibility to have maser emission 
for the curvature radiation process, but we found that it is not possible.
The reason is in the different energy dependence of the single particle emitted power $P$
upon the particle energy $\gamma$: $P\propto \gamma^2$ for synchrotron and $P\propto \gamma^4$
for curvature radiation.
We are thus back to the particle bunching possibility.

The fact that the phenomenon of FRB is observed in radio wavelengths 
with such a short duration suggests that the
particles emit by a primary (non reprocessed) non-thermal process.
The observed power (if extragalactic) suggests a compact object, 
such as a neutron star or an accreting stellar black hole, at the
origin of the energetics.
The large luminosities and energetics imply that the number 
of emitting particles is very large. 
This in turn suggests that the FRB emission comes from the vicinity of the powerhouse
responsible for the injection and acceleration of the emitting particles.
The compactness of the object, the energetics, and the requirement of being close
to the powerhouse suggest a strong value of the magnetic field.
This in turn excludes the synchrotron mechanism, which would produce frequencies that are too large.
The other remaining possibility is curvature radiation from bunches of particles
emitting coherently.
In this prospective the roughly millisecond 
duration gives an even stronger limit on the size of the
emitting region, which is required to be $R< ct\sim$300 km. 
This strengthens the arguments in favour of a compact object as central engine. 

The curvature radiation process has already been discussed by several authors, mainly to explain
the origin of the pulses in pulsars, which also show huge $T_{\rm B}$, and shorter 
duration than in FRBs, but much lower luminosities.
The most recent and complete work suggesting coherent curvature radiation to explain FRBs is
Kumar, Lu, \& Bhattacharya (2017, hereafter K17).
They found a set of constraints limiting the geometry, density, and energy of the emitting particles
in models that can successfully explain FRBs.
Perhaps the most severe limitation they found is about the requirement
on the magnetic field. It has to be equal or stronger than the critical magnetic field
$B_{\rm c}\equiv m_{\rm e}^2 c^3/(he) = 4.4 \times 10^{13}$ G to be comparable with the
Poynting vector of the produced radiation.
This suggests strongly magnetized neutron stars as progenitors.

In this paper, following the study of K17, we consider what limits are posed by the process
of self--absorption, finding solutions aiming to minimize the required total energy.
Furthermore, we consider power law particle distributions as well as mono-energetic particle distributions,
illustrating the effects that coherence has on the observable spectrum, both in the
thick and thin part.
Finally, we study the importance of inverse Compton radiation. 
We illustrate that in the highly ordered geometries, associated with 
curvature radiation in highly magnetized neutron stars, inverse Compton emission
is severely limited.
This agrees with the non-detection, so far, of FRBs in any other frequency band other than the radio
(see Sholtz et al. 2017 and Zhang \& Zhang 2017 for FRB 121102) 
This may imply that the FRB is mainly a {\it radio} phenomenon 
with no counterparts at other frequencies.

\section{Set up of the model}

First let us introduce the notation, that is
\begin{eqnarray}
\rho\, &=&\, {\rm curvature \, radius;}\nonumber \\
\nu_0 &\equiv& \frac{c}{2\pi \rho }; \quad \nu_{\rm c} \equiv  \frac{3}{ 2}  \gamma^3 \nu_0; \nonumber \\
x&\equiv&  \frac{\nu}{\nu_{\rm c}}\equiv \frac{ 2\nu}{3 \nu_0 \gamma^3 }     
\label{def}
,\end{eqnarray}
where $\nu_0$ would correspond to the fundamental frequency if the trajectory 
of the particle were a real circle of radius $\rho$.
The observer sees radiation from the relativistic particles for a time
$(\rho/c)/ \gamma$ of the trajectory and the Doppler effect shortens it by another
factor $\sim 1/\gamma^2$. 
Thus the typical observed frequency $\nu_{\rm c}$ is a factor $\gamma^3$ higher than $\nu_0$. 

The monochromatic power emitted by the single particle is written as (see e.g. Jackson 1999; LG17) 
\begin{equation}
p_{\rm c}(\nu,\gamma)\, =\,   \sqrt{3} \frac{e^2}{ \rho}   
\gamma x \int_x^\infty K_{5/3}(x^\prime) dx^\prime.
\label{psingle}
\end{equation}
Usually, the monochromatic luminosity, which we denote $L(\nu)$, 
is the single electron power multiplied
by the particle density at a given $\gamma$, integrated
over all $\gamma$, and then multiplied by the volume.
This assumes that particles emit incoherently and that they are
distributed isotropically. 
In these conditions as well the emitted luminosity is isotropic.

For ordered magnetic geometry, we must take into account that the observed times are  
Doppler contracted for lines of sight along the particle velocity direction.
Furthermore, the radiation is collimated.

If an observer located orthogonally to the motion measures
a time $\Delta t_{\rm \perp}$, the observer located 
within an angle $1/\gamma$ along the velocity direction receives photons within a time $\Delta t_{\rm \parallel}$, that is
\begin{equation}
\Delta t_{\rm \parallel} \, \sim \,  \frac{ \Delta t_{\rm \perp}}{  \gamma^2} 
\label{ta}
.\end{equation}
As discussed in K17, the volume of a bunch of  particles
observed to emit coherently may extend in the lateral dimensions (lateral with respect to
the velocity direction, which coincides with the line of sight).
On the other hand, extending the lateral dimensions too much requires 
to consider particles moving along magnetic field lines of different curvature radii,  therefore 
emitting at other frequencies, possibly in a incoherent way. 
We assume that along the line of sight the coherence length is $\rho/\gamma$, 
the same length for which the produced radiation reaches us. 
This size is along a single magnetic field line, perpendicular to the radial direction;
radial here means with respect to the centre of the neutron star.
Along the latter, we assume that the relevant size is the same, $\rho/\gamma$.
Perpendicular to this, along the longitudinal direction, the magnetic field lines
could have the same curvature radius for an extension of
the order of $\rho$, if there are indeed particles along the corresponding field lines.
In general we parametrize this uncertainty defining the volume as
\begin{equation}
V \, =\, \frac{\rho^3}{ \gamma^{2+a}}
\label{v}
,\end{equation}
where $a$ can be zero or one.

The observed emission is not isotropic, but it is beamed within a solid angle
much smaller than $4\pi$. 
The single particle  emits mainly within $\Delta \Omega  \sim \pi/\gamma^2$,
but the general case, for an ensemble of particles, depends on the assumed geometry.

As said above for the volume, particles in the longitudinal size may indeed
contribute to the coherent luminosity. 
If so, the corresponding solid angle is $ \pi \rho^2/(\gamma R^2)\sim \pi/\gamma$.
Here $R$ is the distance from the centre of the neutron star.
The other case corresponds to  fewer lines occupied by the emitting particles.
For simplicity, we assume in this case a solid angle $ \pi \rho^2/(\gamma^2 R^2)\sim \pi/\gamma^2$.
Again, we may parametrize the two cases assuming
\begin{equation}
\Delta \Omega \, =\, \frac{\pi}{ \gamma^{1+a}}
\label{domega}
,\end{equation}
where $a$ is the same parameter of Eq. \ref{v}, so that
\begin{equation}
\frac{V}{ \Delta \Omega }\, =\,  \frac{\rho^3}{ \pi   \gamma}, \qquad 
\frac{V^2}{\Delta \Omega }\, =\,  \frac{\rho^6 }{ \pi   \gamma^{3+a}}  
\label{vdomega}
.\end{equation}

Coherent radiation depends on the square of the number of particles contained in the
coherence volume\footnote{The volume for which the particles are observed to emit coherently.
As discussed in K17, in general this is a function of the distance from the sources and 
observer: the longer the distance, the larger the coherence volume can be.}.
Taking into account the Doppler contraction of times and the collimation of the radiation
we have
\begin{eqnarray}
L^{\rm thin}_{\rm iso}(\nu)\, &=&\, \int_{\gamma_1}^{\gamma_2} p(\nu,\gamma) \left[N(\gamma) V\right]^2
\frac{\Delta t_{\rm \perp} }{ \Delta t_{\rm \parallel} } \, \frac{4\pi }{ \Delta \Omega} d\gamma 
\nonumber \\
&=&   
\int_{\gamma_1}^{\gamma_2} 4\, p(\nu,\gamma) \frac{\rho^6}{\gamma^{1+\alpha} } N^2(\gamma) d\gamma 
\label{power}
.\end{eqnarray}

\subsection{Monoenergetic particle distribution}

For a monoenergetic particle distribution we have
\begin{equation}
N(\gamma)\, =\, N_0\delta(\gamma-\gamma_0)
\label{ngamma0}
.\end{equation}
In this case we obtain
\begin{equation}
L^{\rm thin}_{\rm iso}(\nu) =  \frac{16\pi }{ \Gamma(1/3) } \left( \frac{2 \pi \rho}{ 3 c } \right)^{1/3}
 \frac{ e^2 \rho^5 N_0^2  }{ \gamma_0^{1+a} } \nu^{1/3} e^{-\nu/\nu_c}
,\end{equation}
where we used the following approximation:
\begin{eqnarray}
F(\nu/\nu_{\rm c}) &\equiv& \frac{\nu}{ \nu_{\rm c}}  \int^\infty_{\nu/\nu_{\rm c}} K_{5/3}(y)dy 
\nonumber \\
&\sim&  \frac{4\pi }{ \sqrt{3}\Gamma(1/3)} \left( \frac{\nu }{ 2\nu_{\rm c}}\right)^{1/3} e^{-\nu/\nu_c}
\end{eqnarray}
as in Ghisellini (2013), applied in that work for the synchrotron process, but also valid here.

\subsection{Power law particle distribution}

Assume now that the particles are distributed as a power law,
\begin{equation}
N(\gamma)\, =\, N_0\gamma^{-n}
\label{ngamma}
\end{equation}
between $\gamma_1$ and $\gamma_2$.
Eq. \ref{power} with this particle distribution gives
\begin{eqnarray}
L^{\rm thin}_{\rm iso}(\nu)\, & = & A(n)\,  e^2 \rho^5 N_0^2 \left(\frac{\nu}{ \nu_0}\right)^{-(2n+a-1)/3} 
e^{-\nu/\nu_{\rm c, max}}
\nonumber \\
              A(n) &=&  \frac{2}{ \sqrt{3}} 3^{(2n+a-1)/3} \frac{2n+a+4}{ 2n+a+2} 
\nonumber \\
             &~&  \Gamma\left( \frac{2n+a+4}{ 6}\right)      \Gamma\left( \frac{2n+a }{ 6}\right)
\label{plthin} 
.\end{eqnarray}

\section{Absorption}

 The absorption could be coherent, but there is a very important difference:
the absorption cross section and hence the absorption coefficient depends on the
mass of the particle.
Therefore even if particles absorb as the square of their number, as if they were
a single charge $Q=N(\gamma) e$,  their equivalent mass is $M=N(\gamma)m$, such that
the absorption coefficient is $\alpha(\nu) \propto Q^2/M = (e^2/m) N(\gamma)$,
as in the incoherent process (see e.g. Cocke \& Pacholczyk 1975).
Therefore the absorption optical depth is the same as in the incoherent process.

The elementary process of absorption of curvature radiation is 
appropriately described by the corresponding cross section, derived by 
Locatelli \& Ghisellini (2017), as follows:
\begin{eqnarray}
\sigma_\nu  &=&\,  \frac{1}{ 2 m \nu^2}  \left(\frac{d \, p(\nu, \gamma)}{ d\gamma}\right)  \nonumber\\
&=&\, \frac{1}{ 2\sqrt{3}}\, \frac{e^2\rho }{ \gamma^6 mc^2} 
\left[ K_{5/3}(x)  -\frac{2}{ 3} \, \frac{  \int_x^\infty K_{5/3}(y) dy }{ x} \right]
\label{sigma}
.\end{eqnarray}
For small arguments $y$ of the Bessel function, 
$K_a(y) \to 2^{a-1}\Gamma(a) y^{-a}$.
Using this approximation up to $y=1$ and setting $K_a(y)=0$ above,
we find that $\sigma_\nu(\gamma) \propto \nu^{-1}$ at low frequencies
\begin{equation}
\sigma_\nu  \approx\,   \, 
\frac{ \sqrt{3} \Gamma(5/3)}{ 2^{4/3}}\, \frac{e^2\rho }{ \gamma^3 mc^2} \,
\left( \frac{\nu }{ \nu_0} \right)^{-1}, \quad \nu\ll\nu_{\rm c}
\label{sigmana}
.\end{equation}
This also agrees with numerical results.

%
%
%
%
%
%
\subsection{Absorption by a monoenergetic particle distribution}

If the particle distribution is monoenergetic, $N(\gamma) = N_0 \delta(\gamma-\gamma_0)$,
the absorption optical depth for a layer of length $\rho/\gamma_0$ is 
\begin{equation}
\tau_\nu  \, =\, \sigma_\nu ( \gamma_0)  \,   N_0 \, \frac{\rho}{ \gamma_0}  
\label{tau0}
.\end{equation}
The self-absorption frequency $\nu_{\rm t}$ is defined setting $\tau_\nu=1$.
Using the low frequency approximation of Eq. \ref{sigmana} we have
\begin{equation}
\nu_{\rm t}\,   \sim \, \nu_0  \, \frac{\sqrt{3} \Gamma(5/3)}{ 2^{4/3} } \,\frac{e^2\rho^2 N_0 }{ m c^2 \gamma_0^4}
.\end{equation}

\subsection{Absorption by a power law particle distribution}

Using the $N(\gamma)$ distribution of Eq. \ref{ngamma}  and
using a ($\gamma$--dependent) layer of length $\rho/\gamma$ we have
\begin{eqnarray}
\tau_\nu   &=&  
\int_{\gamma_1}^{\gamma_2} \sigma_\nu ( \gamma)  \,   N(\gamma) \, \frac{dR}{ d\gamma} d\gamma
\, =\, - \int_{\gamma_1}^{\gamma_2} \sigma_\nu ( \gamma)  \,   N(\gamma) \, \frac{\rho }{\gamma^2} d\gamma
\nonumber \\
 & = & \frac{ e^2 \rho^2 N_0 }{ 16 \sqrt{3} mc^2 } \, \left( \frac{\nu}{ \nu_0}\right)^{-(n+7)/3} F(n)
\nonumber \\
F(n)  &=&  3^{ (n+1)/3 }\,   \frac{ (n+6)(n+2)}{ n+4 }    \,
 \Gamma \left(\frac{n+6}{ 6}\right) \, \Gamma\left(\frac{n+2}{ 6}\right) 
\label{tau}
,\end{eqnarray}
where $dR/d\gamma= -\rho/\gamma^2$.
We note that the choice of $\rho/\gamma$ as the appropriate length introduces an extra dependence
on the observed frequency with respect to the choice of a fixed length.
The self absorption frequency is
\begin{equation}
\nu_{\rm t}  = \nu_0 \left[ \frac{e^2 \rho^2 N_0 }{ 16 \sqrt{3} mc^2}F(n) \right]^{\, 3/(n+7)} 
.\end{equation}

\section{SED of coherent curvature radiation}

Rather generally, we can account for both the thin and thick part of the spectrum setting, that is
\begin{equation}
L_{\rm iso}(\nu)  =\, L^{\rm thin}_{\rm iso}(\nu)\,\, \frac{ 1-e^{-\tau_\nu} }{ \tau_\nu } 
\label{Liso}
.\end{equation}
With this equation we can calculate the entire spectrum.
The model parameters are
\begin{itemize}
\item The curvature radius $\rho$. 
Since we are dealing with neutron stars, $\rho$ is likely associated with
(be a multiple of) the radius of the neutron star.
The K17 work found a lower limit to the magnetic field that is able to 
guide  the emitting particles, and found that the guiding
magnetic field must be stronger than $B=10^{13}$ G.
If true, this implies that the emitting particles are not too far from the surface
of the neutron star.
For illustration, we use $\rho=10^6$ cm.

\item The particle density $\sim N_0$.
This is to be found considering two limits:
the first is the luminosity produced by the ensemble of the emitting
particles in a coherent way; the second is that self--absorption
can decrease the observed luminosity in a severe way.

\item The particle distribution, assumed to be a power law of index $n$ or monoenergetic. 
We investigate both cases and for the slope $n$
of the power law we preferentially use $n=2.5$, but we  show
how the spectral energy distribution (SED) changes for different $n$.
We consider that the coherent nature implies that the observed radiation, in 
the thin part, is proportional to the square of particle distribution, giving
$L(\nu) \propto \nu^{-(2n+a-1)/3}$ (Eq. \ref{plthin}). 
For a given $n$, the observed spectrum is {\it steeper} with respect to the
incoherent case.

\item The minimum and maximum particle energy, or equivalently, the
corresponding Lorenz factors $\gamma_{\rm min}$ and $\gamma_{\rm max}$.
For simplicity, when treating the power law case, we assume $\gamma_{\rm min}=1$.
The high energy end of the particle distribution is not energetically important 
if $L(\nu)\propto \nu^{-\alpha}$, where $\alpha>1$. 

\item The collimation factor, parametrized by $a$ in Eq. \ref{domega}, associated
with the choice of the emitting volume as in Eq. \ref{v}.
We do not have arguments in favour or against either possibility, so we 
consider both cases.

\item The number of emitting leptons with respect to protons. 
Protons suffer less self-absorption, since $\tau_\nu \propto m^{-1}$.
This makes the solutions with emitting protons more economical.
On the other hand, the regions close to the surface of a highly magnetized
neutron stars are the kingdom of electron--positron pairs, that can greatly outnumber protons.
In any case, we consider both cases.

\end{itemize}

%
\begin{figure} 
\vskip -2.2   cm
\hskip -0.4 cm
\includegraphics[width=8.2cm]{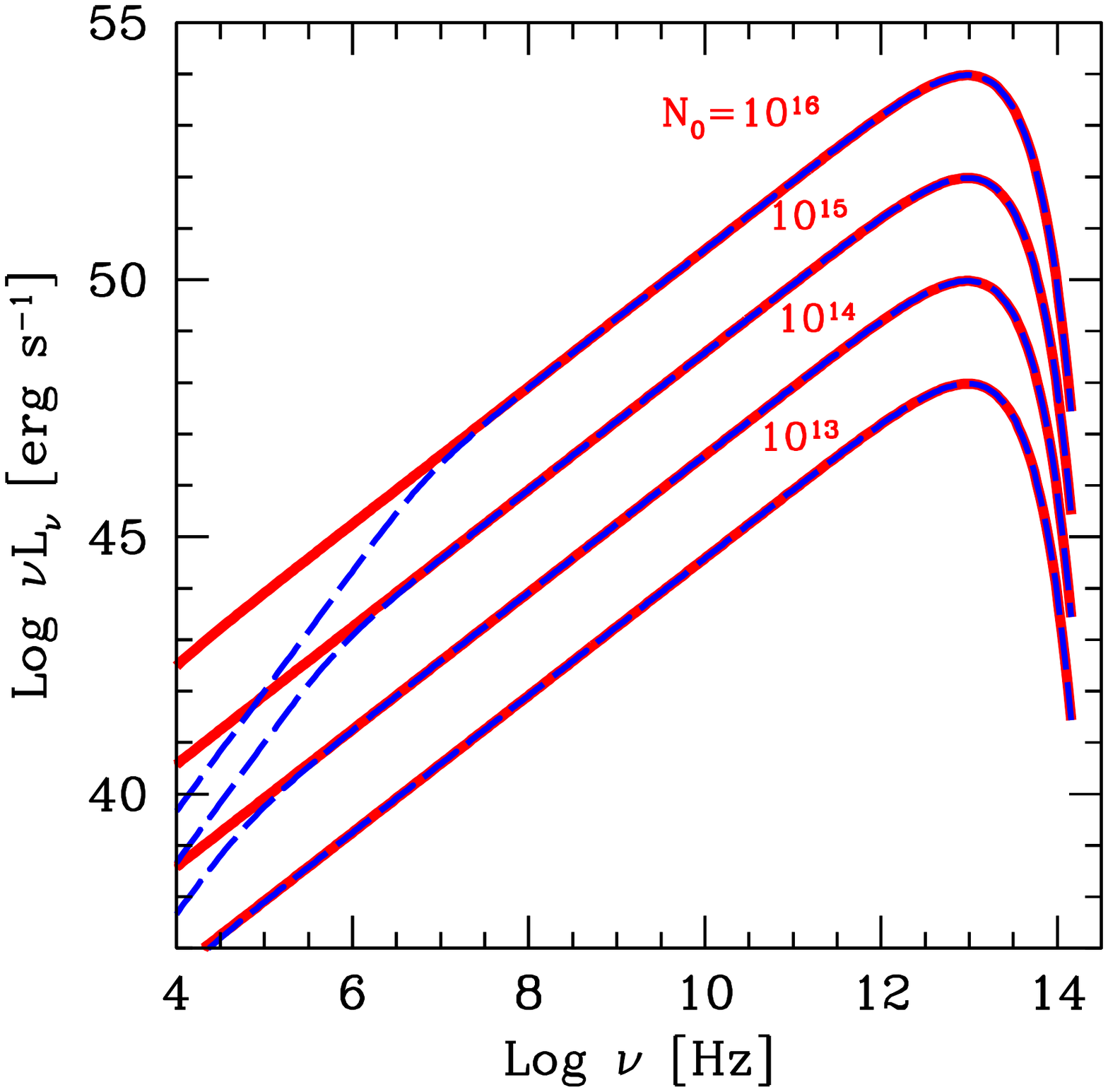}  \\
\vskip -5  cm
\hskip -0.4 cm
\includegraphics[width=8.2cm]{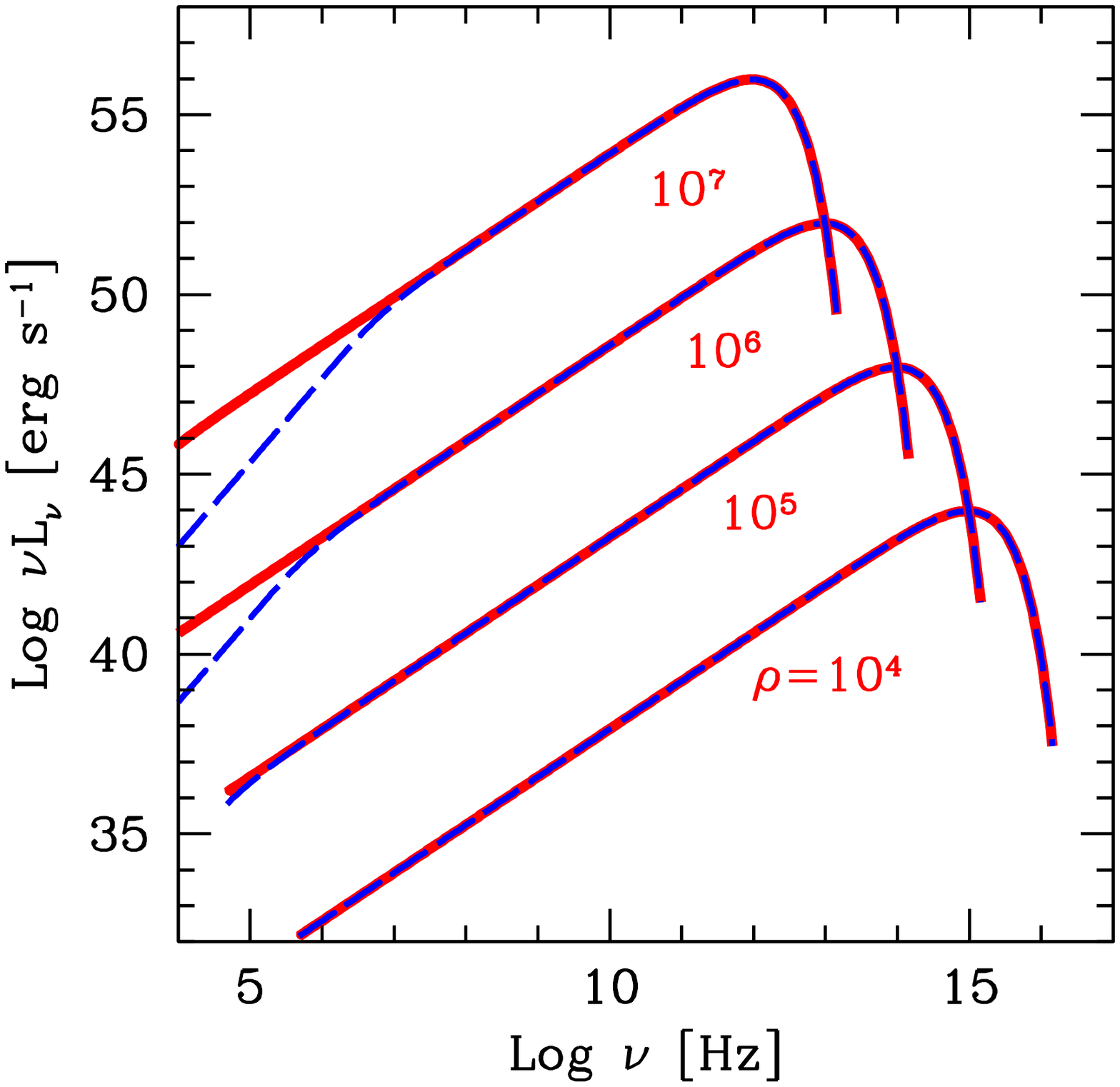} \\ 
\vskip -3.  cm
\hskip -0.4 cm
\includegraphics[width=8.2cm]{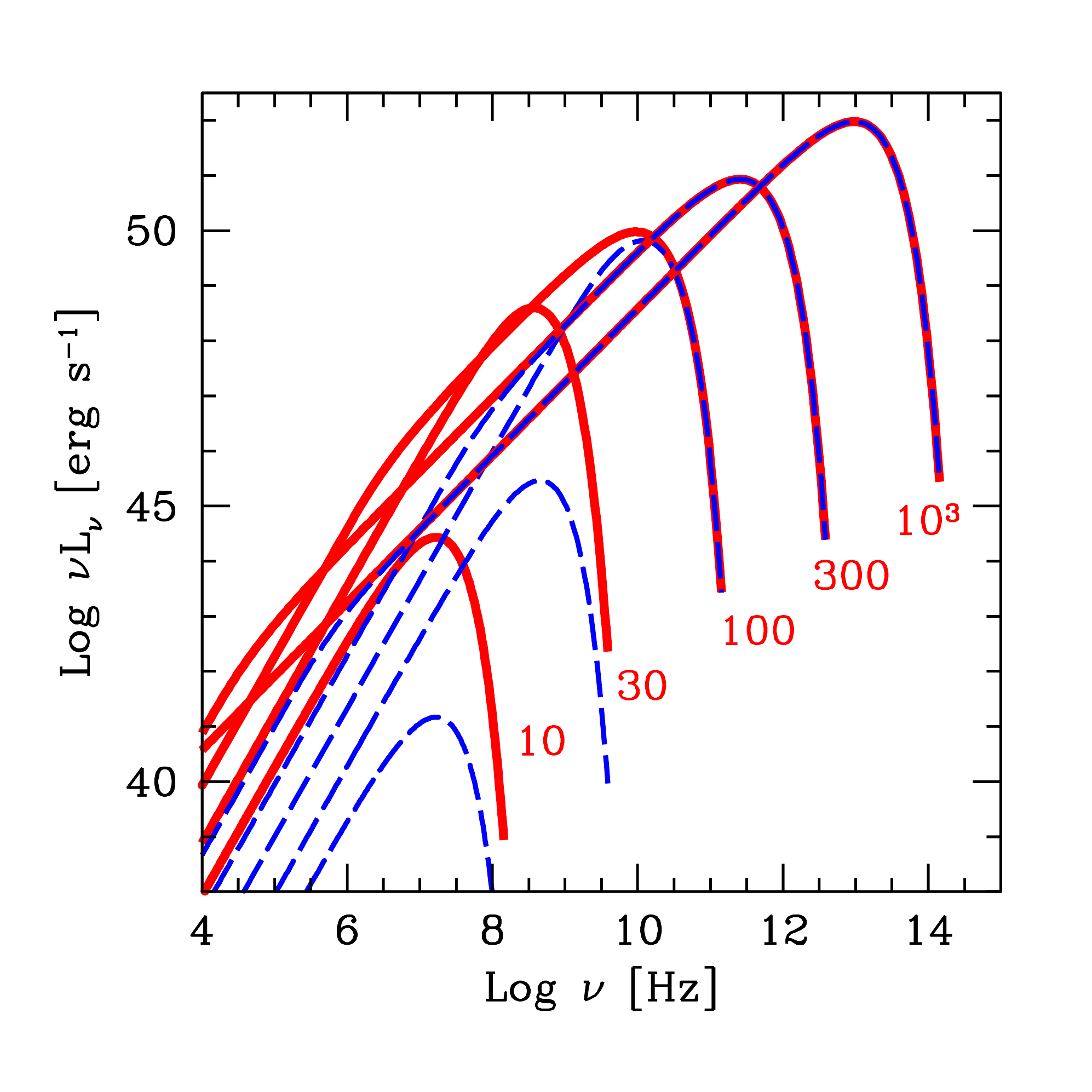} 
\vskip -0.5 cm
\caption{
Examples of SED produced by a monoenergetic particle distribution for changing input parameters.
The reference values of the parameters are $a=1$, $\rho=10^6$ cm, $N_0=10^{15}$ cm$^{-3}$, $\gamma_{\rm max}=10^3$.
The solid lines indicate SED produced by protons; the dashed line indicates SED produced by leptons.
Top panel: The changing density $N_0$ of the particles distribution is shown.
Middle panel: The changing $\rho$ is indicated.
Bottom panel: The changing maximum electron energy $\gamma_{\rm max}$ is shown.
} 
\label{sedparamono}
\end{figure}
\begin{figure*} 
\vskip -2.6 cm
\hskip -0.4 cm
\includegraphics[width=9cm]{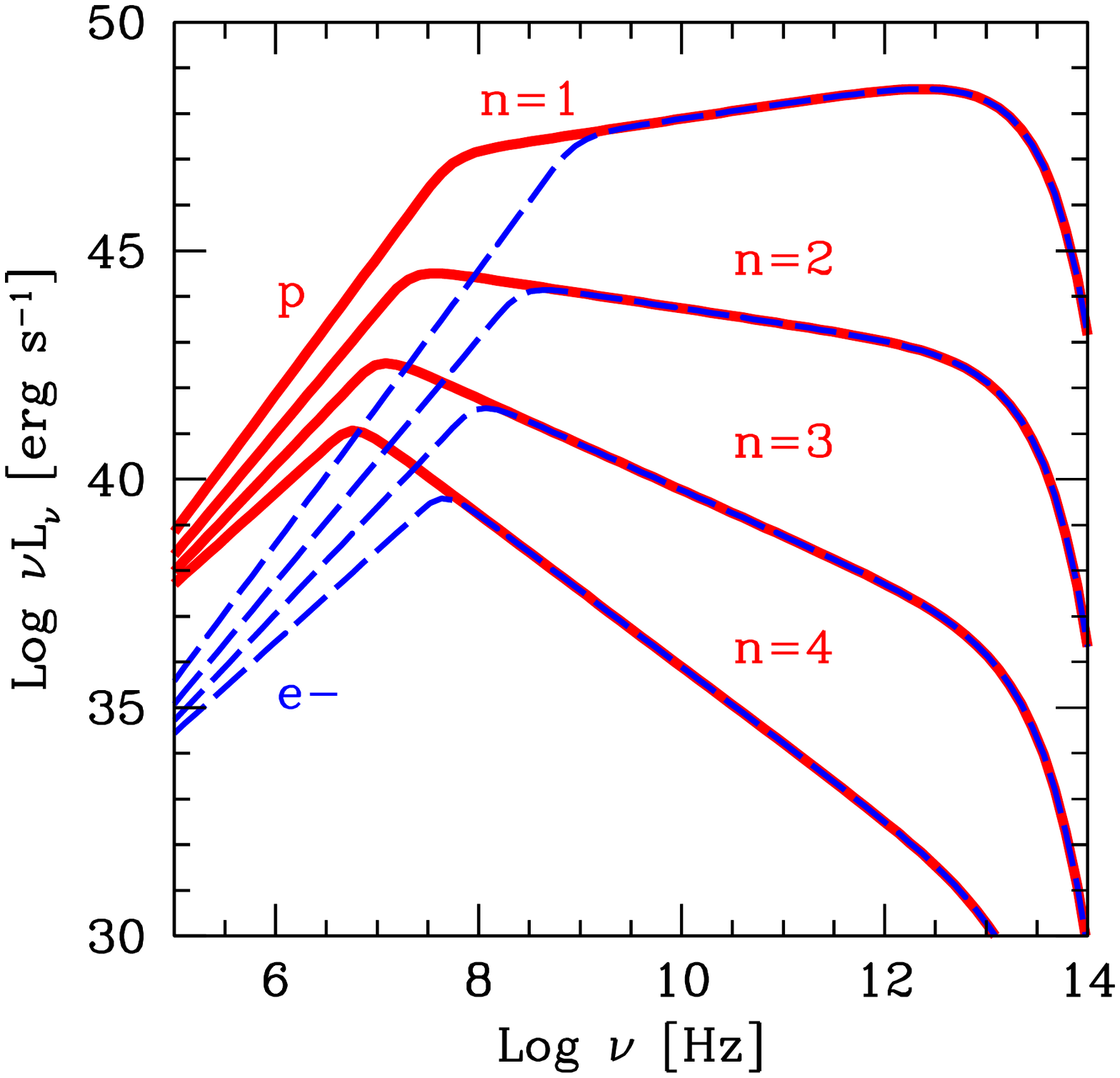} 
\includegraphics[width=9cm]{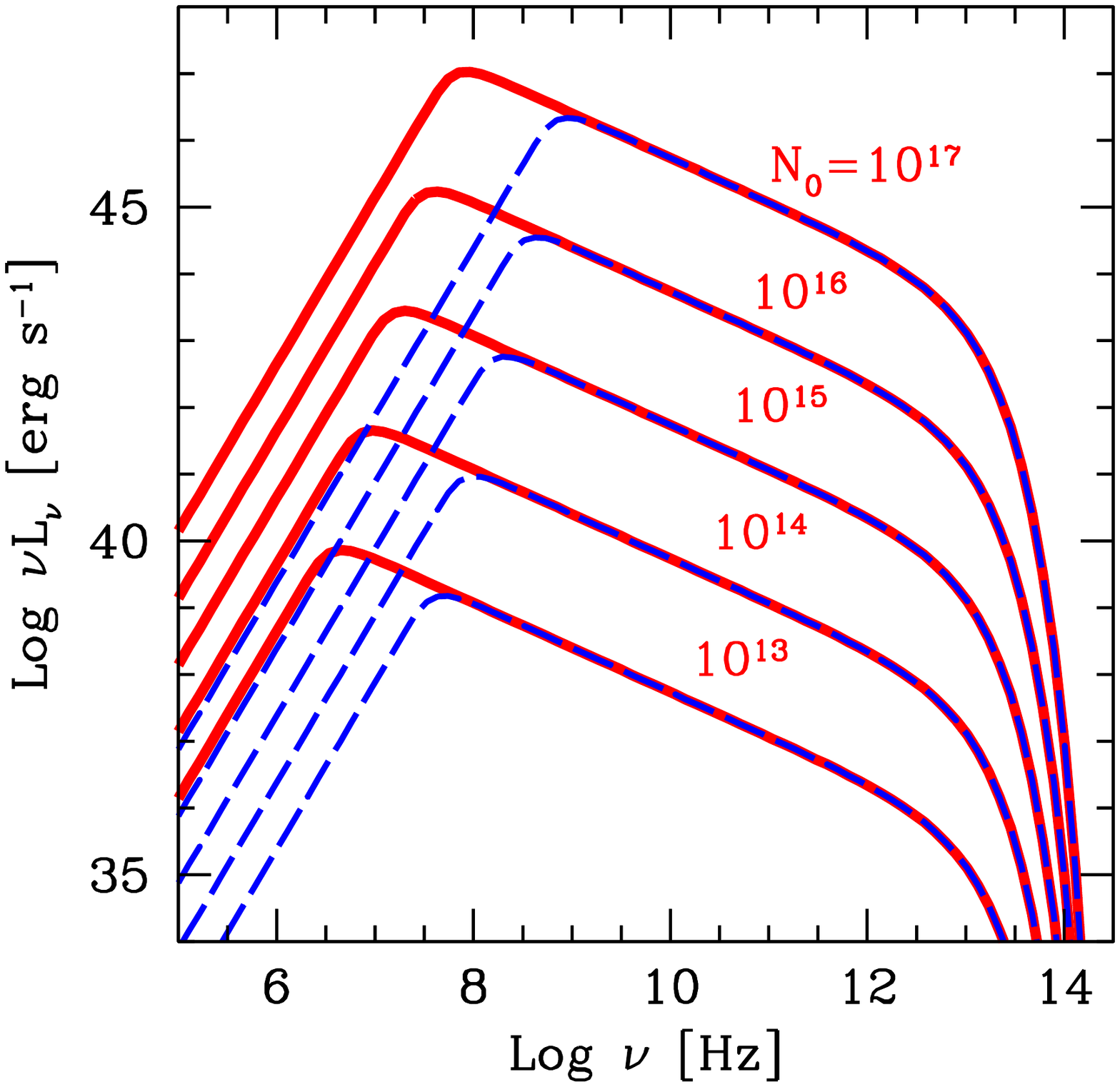}  

\vskip -5 cm
\hskip -0.4 cm
\includegraphics[width=9cm]{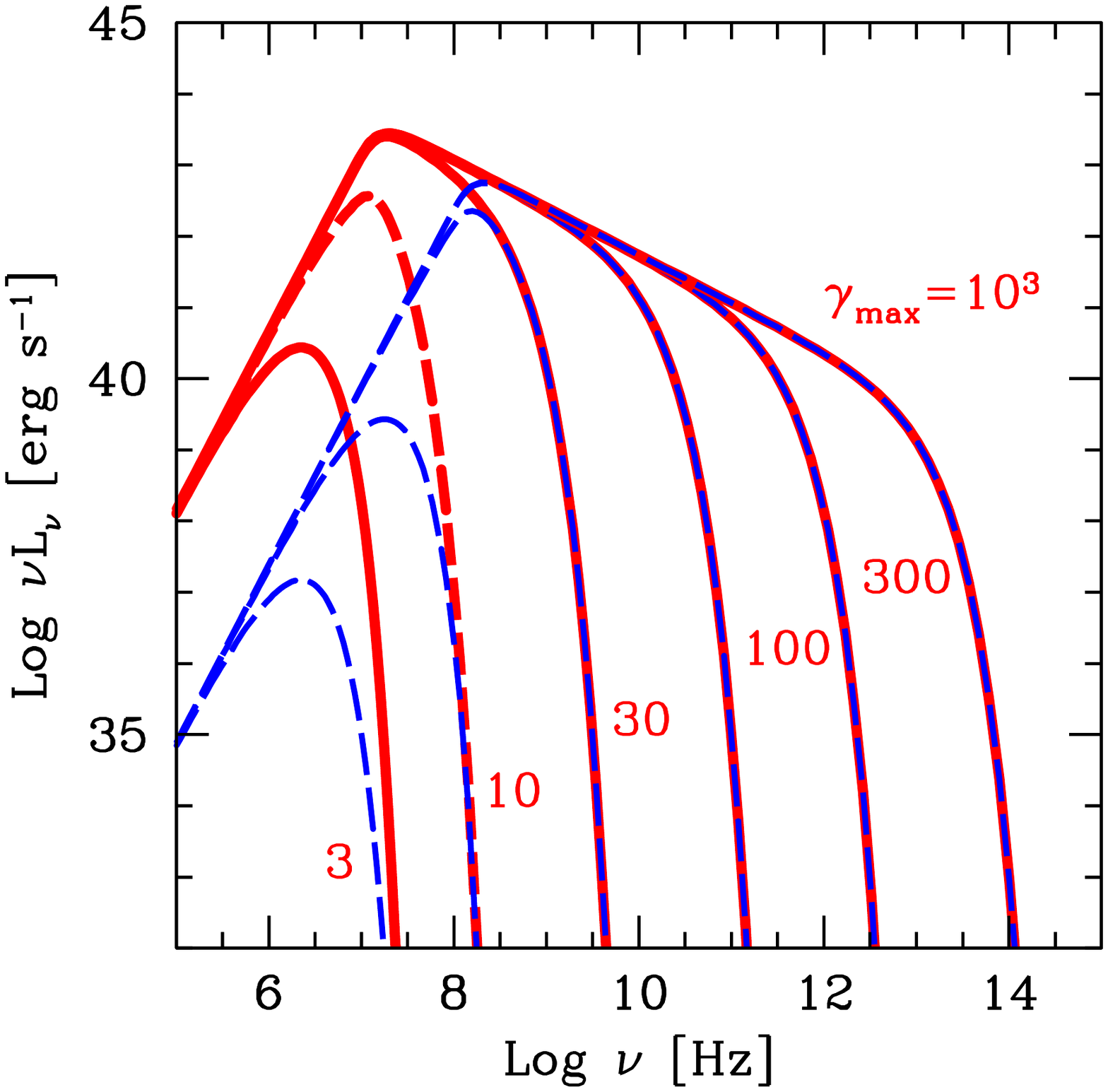} 
\includegraphics[width=9cm]{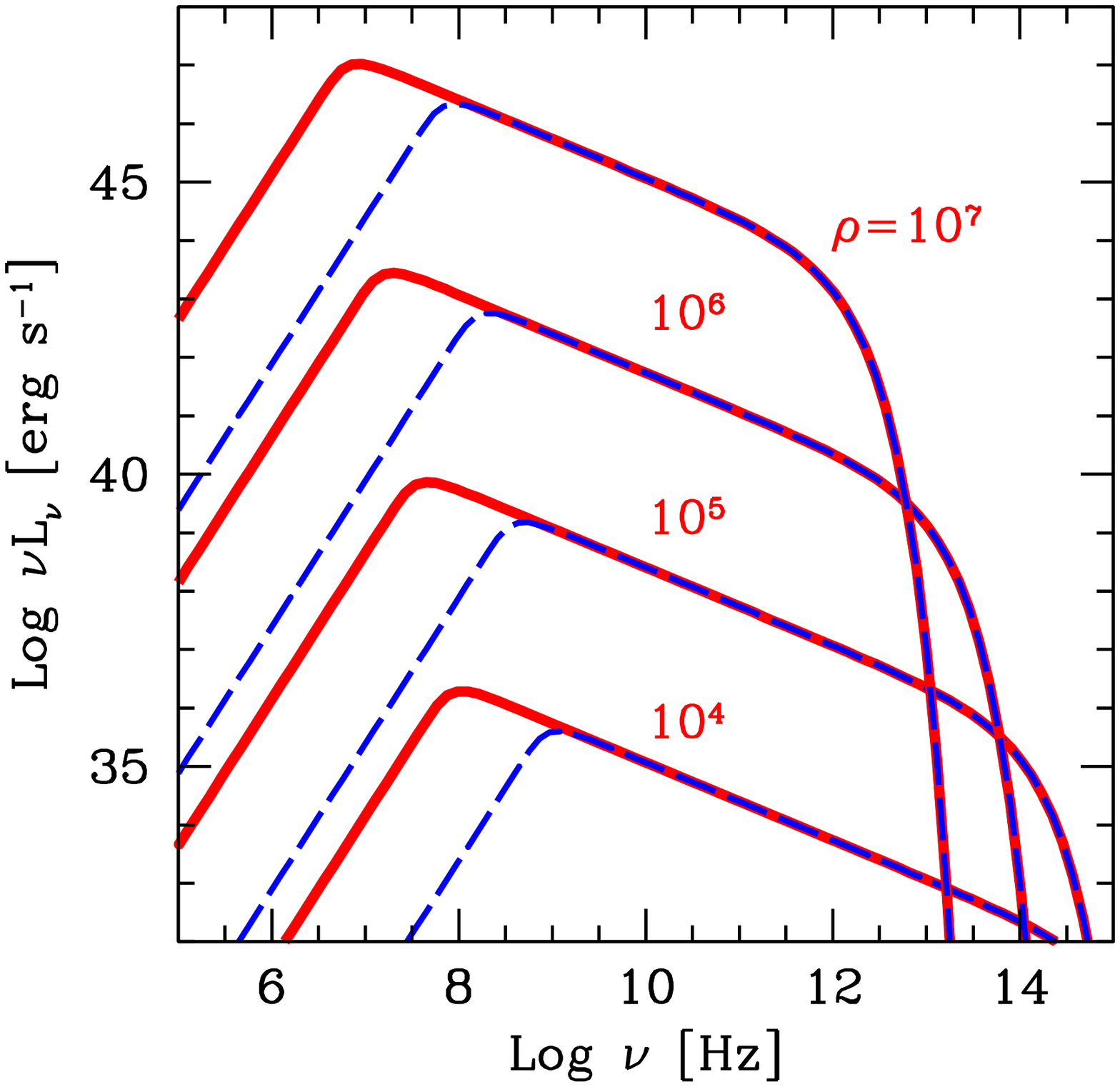} 
\vskip -2.5 cm
\caption{
How the SED changes when changing the input parameters. 
The reference values of the parameters are $a=1$, $\rho=10^6$ cm, $N_0=10^{15}$ cm$^{-3}$,
$n=2.5$, $\gamma_{\rm max}=10^3$.
The solid lines indicate SED produced by protons; the dashed line represents SED produced by leptons.
Top left:
The changing $n$ is shown; contrary to the incoherent case, the self--absorbed spectrum
also changes slope when changing $n$.
Top right: The changing the density $N_0$ of the particles distribution is shown.
Bottom left: The changing the maximum electron energy $\gamma_{\rm max}$ is indicated.
Bottom right: The changing $\rho$ is represented.
} 
\label{sedpara}
\end{figure*}

\subsection{Mononergetic particle distribution}

Fig. \ref{sedparamono} shows how the SED changes by changing one parameter
at a time in the case of a monoenergetic particle distribution.
The reference values of the input parameters are indicated as $\rho=10^6$ cm, $N_0=10^{15}$ cm$^{-3}$, $a=1,$ 
and $\gamma_{\rm max}=10^3$.
For all the figures shown in this paper, we chose the $\nu-\nu L_\nu$ representation.
In this way we see in which band most of the luminosity is produced.
The $\nu L_\nu$ luminosity is a proxy for the bolometric luminosity.

\vskip 0.2 cm
\noindent
{\it Changing particle density ---}
The top panels shows the SED produced by protons (solid line) and
leptons (dashed line) for different densities. 
We note that the absorption for leptons occurs at greater frequencies.

\vskip 0.2 cm
\noindent
{\it Changing $\rho$ ---} 
The middle panel shows SEDs produced by different curvature radii  $\rho$. 
The emission is stronger for greater $\rho$.
In this case the emitting volume is larger, which more than compensates for 
the smaller acceleration.

\vskip 0.2 cm
\noindent
{\it Changing $\gamma_{\rm max}$ ---}
The bottom panel shows that for small values of $\gamma_{\rm max}$ the
emission is completely self--absorbed, and remains so for larger values
of $\gamma_{\rm max}$ for leptons than for protons.
Once the optically thin regime is achieved, we observe the $\nu^{1/3}$ slope.

\subsection{Power law particle distribution}

We show in Fig. \ref{sedpara} how the SED changes by changing one parameter
at a time, keeping the others fixed.
The reference values of the input parameters are $\rho=10^6$ cm, $N_0=10^{15}$ cm$^{-3}$, 
$n=2.5$, $a=1,$ and $\gamma_{\rm max}=10^3$.

\vskip 0.2 cm
\noindent
{\it Changing slope ---}
The top left panel of Fig. \ref{sedpara} shows SEDs with 
different slopes $n$ of the power law distribution of particles.
The self-absorbed spectrum 
$\propto L_{\rm iso}^{\rm thin}(\nu)/\tau_\nu \propto \nu^{7/3}\nu^{-(n+a-1)/3}$.
The slope does depend on $n$, contrary to the incoherent case, for which the self--absorbed spectrum
is $\propto \nu^2$.
Since the absorption coefficient is smaller for protons, the corresponding self-absorbed spectrum
has a higher luminosity than for leptons, and the self-absorption frequency is smaller.

\vskip 0.2 cm
\noindent
{\it Changing particle density ---}
The top right panel of Fig. \ref{sedpara} shows the  SEDs for
different density $N_0$ of the power law distribution of particles.
We note that the thick part of the spectrum $\propto N_0$, while the thin part $\propto N_0^2$.

\vskip 0.2 cm
\noindent
{\it Changing $\gamma_{\rm max}$ ---}
The bottom left panel of Fig. \ref{sedpara} shows the SEDs for 
different $\gamma_{\rm max}$.
For small $\gamma_{\rm max}$, the spectrum is entirely self-absorbed, peaking at the 
characteristic maximum frequency $\nu_{\rm c}$. 
Larger $\gamma_{\rm max}$  allows the thin part of the SED to be visible, that
is equal for protons and leptons.

\vskip 0.2 cm
\noindent
{\it Changing $\rho$ ---}
The bottom right panel of 
Fig. \ref{sedpara} shows the SEDs when changing $\rho$.
We note that the maximum frequency $\propto \nu_0 \propto \rho^{-1}$.

\begin{table*} 
\centering
\begin{tabular}{lllll lllll lll}
\hline
\hline
Figure  &$a$ &$n$ &$\rho$ &$N_0$  &$\gamma_{\rm min}$ &$\gamma_{\rm max}$ &$\log\nu_{\rm t, e}$ 
&$\log\nu_{\rm t, p}$ &$\log\nu_{\rm c}$  &$\log L_{\rm e, max}$ &$\log L_{\rm p, max}$\\
\hline 
%
Fig. \ref{sedpl}   &0  &2.5  &1e6 &3.0e14 &1 &500   &8.1            &7.0        &12.86  &43.23 &43.57   \\
Fig. \ref{sedpl}   &1  &2.5  &1e6 &2.0e15 &1 &500   &8.3            &7.3        &12.86  &43.30 &43.95   \\
Fig. \ref{sedpl}   &0  &2.5  &1e7 &4.2e12 &1 &500   &7.2            &6.1        &11.86  &43.51 &43.85   \\
Fig. \ref{sedpl}   &1  &2.5  &1e7 &4.2e13 &1 &500   &7.4            &6.4        &11.86  &43.80 &44.56   \\
\hline
\hline 
\end{tabular}
\vskip 0.2 true cm
\caption{
Parameters used for the models shown in 
Fig. \ref{sedpl}.
}
\label{para}
\end{table*}

\section{Application to fast radio bursts}

We are looking for models that are able to reproduce the typical characteristic 
observed properties of FRBs, namely $\nu L_\nu\sim 10^{43}$ erg s$^{-1}$ 
at $\nu\sim$1 GHz, and that are compatible in size with the observed duration of $\sim$1 ms.
If these properties can be reproduced, then the model is automatically consistent
with the observed brightness temperatures.
Since the information about {\it the slope} of the radio spectrum is uncertain,
we do not constrain it; this means that, in principle, we allow for both
optically thin and optically thick cases.
Another criterion we adopt is to look for models requiring the minimum amount
of energy.
\begin{figure} 
\vskip -0.5 cm
\hskip -0.7 cm
\includegraphics[width=9.5cm]{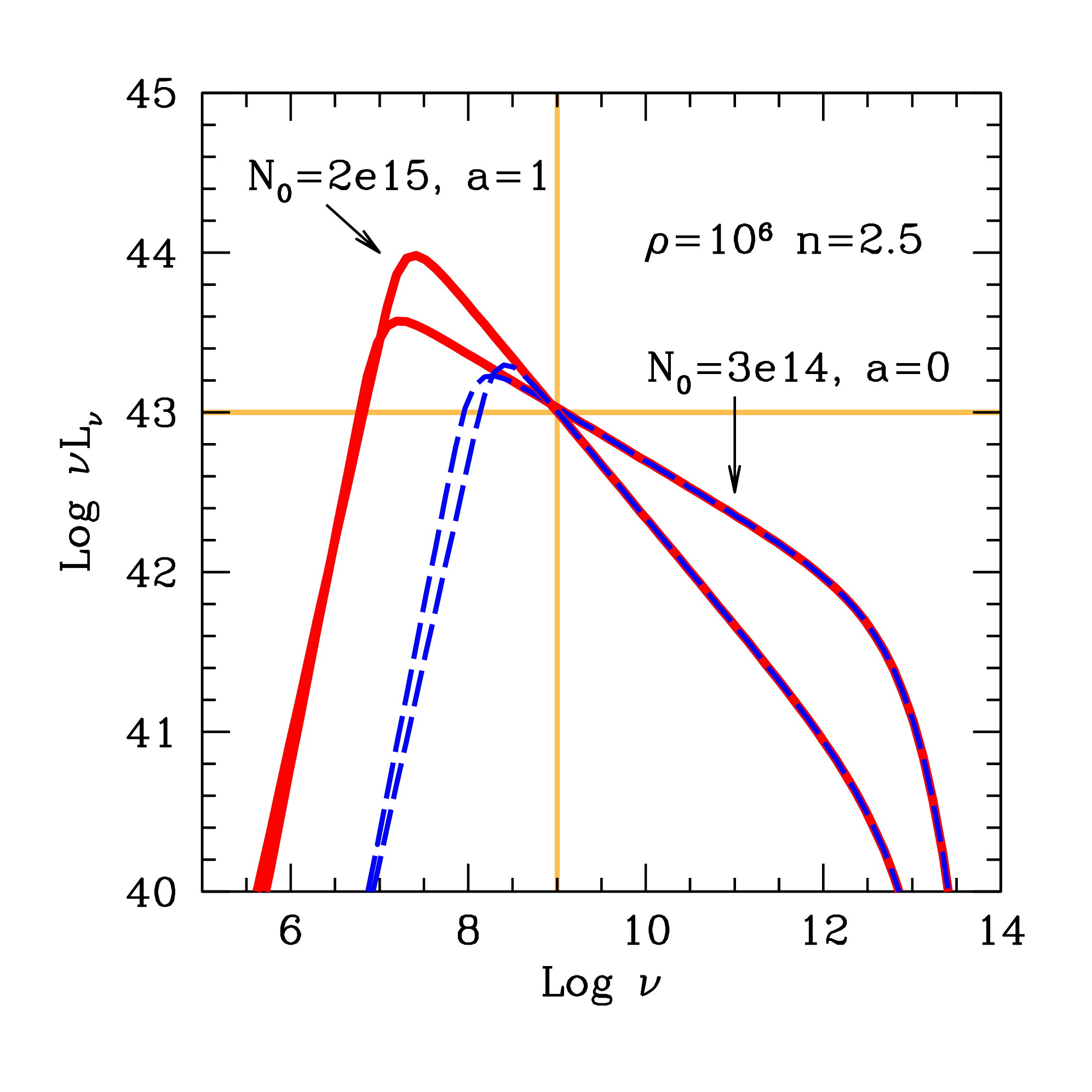} 

\vskip -1.2 cm
\includegraphics[width=9.5cm]{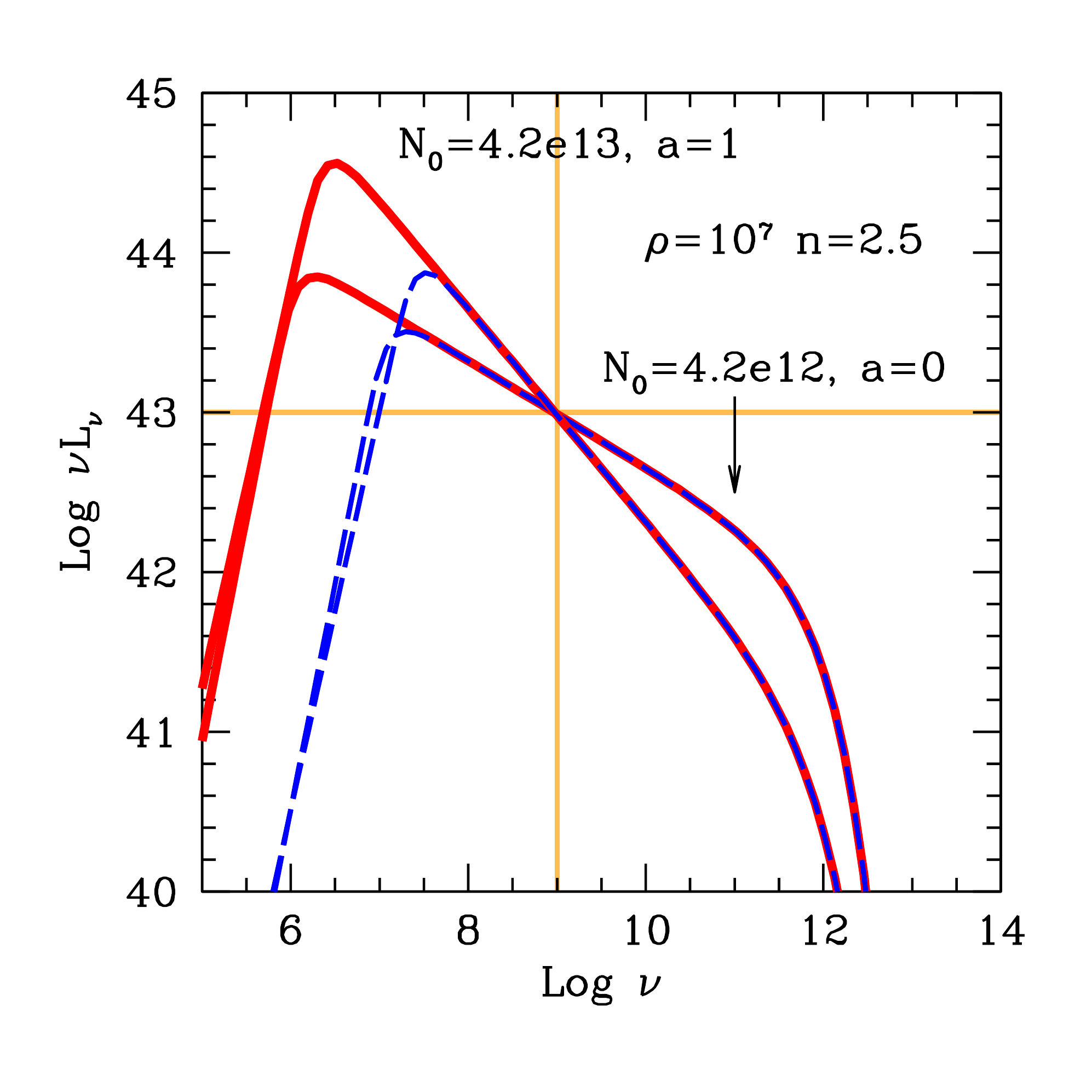} 
\vskip -0.6 cm
\caption{
Spectra, in the $\nu$--$\nu L_\nu$ representation, produced by a power law distribution of particles
accounting for a luminosity of $10^{43}$ erg s$^{-1}$ at 1 GHz.
The adopted slope is $n=2.5$.
The solid lines show spectra produced by protons; the dashed lines represent the
spectra produced by leptons. 
The different slopes are due to the different $a$ value, controlling
the amount of collimation of the radiation, and not to a different
slope of the electron distribution.
The top panel gives examples for $\rho=10^6$ cm.
The bottom panel shows examples for $\rho=10^7$ cm.
Parameters, listed in Tab. \ref{para}, 
were found to minimize the required total energy.
$N_0$ indicated in units of cm$^{-3}$ and $\rho$ in cm.
} 
\label{sedpl}
\end{figure}

\subsection{Illustrative examples}

Table \ref{para} reports the chosen values of the parameters.
In general, the cases with $a=0$ are more economical, since the 
optically thin luminosity depends on $V^2/ \Delta \Omega \propto \gamma^{-3-a}$
(Eq. \ref{vdomega}).
In these illustrative examples, the SED is always optically thin at 1 GHz
and the slope $\alpha \equiv (2n+a-1)/3$ of $L(\nu)\propto \nu^{-\alpha}$  
is $\alpha=5/3$ ($a=1$) or $\alpha=4/3$ ($a=0)$.
In the self--absorbed regime ($\tau_\nu\gg 1$), we have $L(\nu) \propto
L_{\rm iso}^{\rm thin} /\tau_\nu \propto \nu^{(8-n-a)/3}$.
As mentioned above, the slope of the self-absorbed part of the SED does depend
on $n$. In our case, with $n=2.5$, we have $L(\nu)\propto \nu^{11/6}$ (if $a=0$)
or $L(\nu)\propto \nu^{3/2}$ (if $a=1$).
Fig. \ref{sedpl} shows two sets of models with a power law particle distribution
with the same slope $n=2.5$  and different $\rho$.
In Fig. \ref{sedpl} the vertical and horizontal lines correspond 
$\nu=1$ GHz and $\nu L_\nu= 10^{43}$ erg s$^{-1}$.

\subsection{Suppression of the inverse Compton process}

Usually, the importance of the inverse Compton process is measured by the 
Comptonization parameter $y$, defined as
\begin{eqnarray}
y &=& { \rm  average\, number\, of \, scattering } \, \nonumber \\
&\times& 
{\rm  \, fractional \, energy\,  gain\,  per\,  scattering}
.\end{eqnarray}
Compton up-scattering is important for $y>1$.
If particles are relativistic and the seed photons
are isotropically distributed, we have 
$y \sim \tau_{\rm T} \langle \gamma^2\rangle$ when
the scattering optical depth $\tau_{\rm T}$ is smaller than unity.
In our case of highly ordered geometry, both the average number of scatterings
and the fractional energy gain of the scattered photons are largely reduced.
In fact, the curvature radiation process produces photons moving along the same
direction of the emitting particle: the typical angle between seed photons 
and particles is of the order of $\psi = 1/\langle \gamma\rangle $.
Since the rate of scatterings is $\propto (1-\beta\cos\psi) \sim (1-\langle\beta^2\rangle)$,
we have a reduction of a factor $1/\langle \gamma^2\rangle$ with respect to an isotropic case.

The frequency of the scattered photons $\nu_1$ depends upon the 
angles between the electron velocity and the photons before ($\psi$) and after
($\psi_1$) the scattering. 
Both are measured in the observer frame.
We have (see e.g. Ghisellini 2013)
\begin{equation}
\nu_1 = \nu \, \frac{1-\beta \cos\psi }{ 1-\beta\cos\psi_1} \approx \nu
.\end{equation}
In the comoving frame, the pattern of the scattered radiation has a backwards--forwards symmetry,
and this implies that in the observed frame, most of the scattered radiation in concentrated along
the electron velocity vector, within an angle $\psi_1\sim 1/\gamma$, independent of $\psi$. 
As a result, not only the scattering rate is largely reduced, but the scattered
photons, on average, do not change their frequency.


Another possible process is inverse Compton scattering with photons 
produced by the hot surface of the neutron star.
For monoenergetic particles with Lorentz factor $\gamma$, 
the observed luminosity, accounting for collimation, is
\begin{equation}
L_{\rm ext} \sim \frac{4\pi }{ \Delta \Omega} V \sigma_{\rm T} c U_{\rm ext} N_0 \gamma^2 
,\end{equation}
where $N_0$ is the approximately the particle density and $U_{\rm ext}$ is the radiation energy density 
produced externally to the scattering site.
If the latter is close to the neutron stars surface of temperature $T$, $U_{\rm ext} \sim aT^4$
In this case we have
\begin{equation}
L_{\rm ext} \sim 4\pi \rho^3 \sigma_{\rm T} c a T^4 n \langle \gamma \rangle
 \sim 1.9\times 10^{32} \rho_6^3 T_6^4  N_{0, 15} \langle \gamma_2\rangle \,\, {\rm erg \, s ^{-1}}
.\end{equation}
This luminosity should last the same time as the FRBs and should be observed
at $h\nu \sim 3\gamma^2 kT \sim$ 2 $\gamma_2^2 T_6$  MeV.   

We conclude that it is very unlikely that the inverse Compton process plays
a significant role in the proposed scenario.
Therefore we expect that FRBs are weak or very weak emitters of X-rays or $\gamma$-rays.

\section{Cooling timescales}

In the absence of coherent effects, 
the frequency integrated power emitted by a single particle is
\begin{equation}
P_{\rm c} = \frac{2e^2\gamma^4 c}{ 3\rho^2}
.\end{equation}
Instead, for coherent emission, the power emitted by the single particle 
depends on the square of the number of other particles emitting in the same volume $V$ 
of the bunch.
The cooling timescale can be calculated 
dividing the total energy of particles by the total power they emit as follows:
\begin{eqnarray}
t_{\rm cool} \,& =&\,  \frac{ [V N(\gamma)] \gamma mc^2 }{ [V N(\gamma)]^2 P_{\rm c} }
 =\,  \frac{  \gamma mc^2}{  V N(\gamma)  P_{\rm c} } \,
 =\,  \frac{ 3 \gamma^{n+a-1}  mc^2 }{  2e^2 N_0 \, \rho\, c }   \nonumber  \\
\, & =&\, 1.7\times 10^{-17} \, \frac{ \gamma^{n+a-1} }{ N_{0, 13}\, \rho_6  }  \,\, {\rm s}
,\end{eqnarray}
where we have assumed $m=m_{\rm e}$, $N_0=10^{15} N_{0, 15}$ cm$^{-3}$ and $\rho=10^6\rho_6$ cm. 
For $n>1$, $t_{\rm cool}$ {\it increases} for larger $\gamma$: the radiative process is
more efficient for {\it lower energy} particles.
This strange behaviour occurs because (for positive $n$) there are 
fewer and fewer particles for increasing $\gamma$, and thus the power of coherent emission decreases.
However, this is true for thin coherent emission.
Low energy particles self-absorb the radiation they themselves produce, inhibiting
their radiative cooling.
In the absence of re-acceleration or injection of new particles, most of the
radiation is then emitted at the self-absorption frequency (by electron of
corresponding Lorentz factor $\gamma_{\rm t}$), which must evolve
(decrease) very quickly.  
The cooling time is therefore long for electron energies $\gamma \ll \gamma_{\rm t}$,
has a minimum at $\gamma\gsim \gamma_{\rm t}$ and then 
increases again for $\gamma\gg \gamma_{\rm t}$.

The typical value of $\gamma$ required to emit 1 GHz emission is 
$\gamma \sim (4\pi\nu_{\rm c}/3c)^{1/3} \sim 52 (\rho_6 \nu_{\rm c, 9})^{1/3}$.
At 1 GHz, for $n=3$, $N_0=10^{13}$ cm$^{-3}$ and $a=1$, $t_{\rm cool}\sim 2.4 \times10^{-12}$ s for any $\rho$.
In this extremely short timescale the particle can travel for a mere $7\times 10^{-2}$ cm.
This severe problem can be relaxed if fewer particles per bunch are needed to produce the observed 
luminosity, namely if the observed emission is produced by many ($M$) bunches, each containing 
$VN_0/M$ particles.

Alternatively, we should have an efficient acceleration mechanism that is able to balance the
radiation losses.
One possibility is an electric field parallel to the magnetic field.
In this case, particles with $\gamma\ll \gamma_{\rm t}$ increase their energy up to
the point where radiative losses dominate. 
Particles would tend to accumulate at the energy where gains and losses balance. 
In this scenario (acceleration of low energy particles, no re-injection,  and cooling)
the emitting particle distribution would be quasi-monoenergetic.

Even if we can conceive some solutions, we consider the problem of these extremely 
short radiative cooling timescales very challenging.
We not only require either many bunches or a re--acceleration
mechanism, but we have still to explain why the typical duration of the observed
pulses is $\sim$1 ms: it cannot be associated with a typical radiative cooling timescale.

\section{Discussion and conclusions}

We studied the coherent curvature radiation process and its self-absorption
process to find if a region of the parameter space exists that is allowed
to reproduce the observed properties.
To this aim, we were forced to consider a well-defined 
geometry with a high degree of order, located close to the surface of a neutron star,
where it is more likely to find the required large densities of relativistic particles
responsible for the coherent emission.
Furthermore, we discussed the main properties of the SED predicted in this case, which is qualitatively different from 
the non-coherent case. 
Owing to Doppler time contraction (Eq. \ref{ta})
the size $\rho/\gamma$ visible to the observer corresponds to a time $\rho/(c\gamma^3)$ 
and to an observed wavelength $\lambda_{\rm c}=c/(\gamma^3\nu_0)$. In other
words, the observed wavelength is short, but particles are distributed in a much larger region and
can still emit coherently.

We did not investigate the reasons for having such a short and powerful burst of energy 
in the vicinity of the neutron star. 
However, the involved energetics are not extreme for a neutron star.
The isotropic equivalent observed energetic is $E_{\rm iso} \sim 10^{40}$ erg, which
can be reduced by a factor $10^2$--$10^3$ if the emission is collimated.
Giant flares from magnetars, by comparison, can reach energetics one million times larger
(see e.g. Palmer et al. 2005 for SGR 1806--20).
This implies that repetition is well possible from the energy point of view.

The findings and conclusions of our work can be summarized as follows:

\begin{itemize}
\item Self-absorption of coherent curvature radiation is an important process towards constraining
the allowed choices of the parameters, such as the curvature radius and the particle density.
On the other hand, there is indeed a region of the parameter space in which the observed
properties of FRBs can be reproduced.

\item The emitted SED is likely to be steep (i.e. $\alpha>1$, for $F(\nu)\propto \nu^{-\alpha})$
above the self-absorption frequency.
Because of the limited range of the allowed values of the parameters, it is likely
that the emission is self-absorbed at frequencies below 1 GHz.
This predicts that the observed radio spectrum at $\sim$1 GHz, once scintillation is accounted for, is
either close to its peak (i.e. $\alpha\sim 0$) or steep ($\alpha>1$), at least when
leptons dominate the emission.

\item Coherent curvature radiation is possible if the geometry of the magnetic field and
the emitting region is well ordered. 
Curvature photons are produced along the same direction of the particle velocities, 
much depressing the possibility to scatter.
Therefore the model predicts an absent or very weak inverse Compton emission
at high energies.

\item This makes the $\sim$GHz-millimeter  band the preferred band where FRBs
can be observed.
The weakness of the produced emission at high (X and $\gamma$--ray) energies 
is a strong  prediction of the model.
\end{itemize}

\section*{Acknowledgements}
We thank the anonymous referee for her/his critical comments
that helped improve the paper.





\begin{thebibliography}{}

\bibitem[]{beloborodov2017} Beloborodov A.M., 2017, ApJ, 843, L26

\bibitem[]{chatterjee2017} Chatterjee S., Law C.J, Wharton R.S. et al. 2017, Nature, 541, 58

\bibitem[]{cocke1975} Cocke W.J. \& Pacholczyk A.G., 1975, ApJ 195, 279

\bibitem[]{cordes2016} Cordes J.M. \& Wasserman I., 2016, MNRAS, 457, 232 

\bibitem[]{falcke2014} Falcke H. \& Rezzolla L., 2014, A\&A, 562, A137

\bibitem[]{fuller2015} Fuller J. \& Ott, C.D., 2015, MNRAS, 450, L71

\bibitem[]{ghisellini1991} Ghisellini G. \& Svensson R., 1991, MNRAS, 252, 313

\bibitem[]{ghisellini2013} Ghisellini G. 2013, {it Radiative processes in high energy astrophysics}, 
        Lecture Notes in Physics, 873, Springer Switzerland
              
\bibitem[]{ghisellini2017} Ghisellini G. 2017, MNRAS, 465, L30 

\bibitem[]{jackson99} Jackson J.D., 1999, Classical electrodynamics, third edition,
John Wiley and Sons, Inc.  

\bibitem[]{kashiyama2013} Kashiyama K., Ioka K. \& Meszaros P., 2013, ApJ, 776, L39

\bibitem[]{katz2017} Katz J.I., 2017, MNRAS, 469, L39 

\bibitem[]{kumar2017} Kumar P., Lu W. \& Bhattacharya M., 2017, MNRAS, 468, 2726 (K17)

\bibitem[]{locatelli2017} Locatelli N. \& Ghisellini G. 2017, submitted to A\&A  (arXiv 1707.06352) (LG17)

\bibitem[]{loeb2014} Loeb A., Shvartzvald Y. \& Maoz D., 2014, MNRAS, 439, L46

\bibitem[]{lingam2017} Lingam M. \& Loeb, A., 2017, ApJ, 837, L23


\bibitem[]{liubarsky2014} Lyubarsky Y., 2014, MNRAS,442, L9 

\bibitem[]{maoz2015} Maoz D., Loeb A. \& Shvartzvald Y et al., 2015, MNRAS, 454, 2183 

\bibitem[]{palmer2005} Palmer D.M., Barthelmy S., Gehrels N. et al., 2005, Nature, 434, 1107 

\bibitem[]{popov2010} Popov  S.B. \& Postnov K.A, 2010,  Proc. of the Conference 
held 15--18 Sep 2008 in Yerevan and Byurakan, Armenia, Editors: H.A. Harutyunian, A.M. 
Mickaelian \&  Y. Terzian, Yerevan, Publishing House of NAS RA, p. 129-132

\bibitem[]{scholtz2017} Scholtz P., Bogdanov S., Hessels J.W.T. et al., 2017, ApJ in press (arXiv:1705.07824)

\bibitem[]{tendulkar2017} Tendulkar S.P., Bassa C.G., Cordes J.M. et al., 2017, ApJ, 834, L7

\bibitem[]{thornton2013} Thornton D., Stappers B., Bailes M., et al. 2013, Science, 341, 53          

\bibitem[]{totani2013} Totani, T. 2013, PASJ, 65, L12

\bibitem[]{zhang2014} Zhang, B. 2014, ApJ, 780, L21

\bibitem[]{} Zhang B.--B. \& Zhang B., 2017, ApJ in press (arXiv:1705.04242)


\end{thebibliography}
\end{document}